\documentclass{ws-procs975x65}

\def\rfr#1{eq.(\ref{#1})}

\def\eqi{\begin{equation}}
\def\eqf{\end{equation}}
\def\eqia{\begin{eqnarray}}
\def\eqfa{\end{eqnarray}}

\def\rp#1#2{{#1\over#2}}

\def\lb#1{\label{#1}}
\def\lg{LAGEOS}
\def\lgg{LAGEOS II}

\begin{document}

\def\nocropmarks{\vskip5pt\phantom{cropmarks}}

\let\trimmarks\nocropmarks

\markboth{Lorenzo Iorio} {Measuring the Lense--Thirring effect
with a LAGEOS-LAGEOS II-OPTIS mission}

\catchline{}{}{}

\title{ON THE POSSIBILITY OF MEASURING THE LENSE--THIRRING EFFECT WITH A LAGEOS-LAGEOS II-OPTIS MISSION}

\author{\footnotesize LORENZO IORIO}

\address{Dipartimento di Fisica dell'Universit$\grave{ a}$ di Bari\\
Via Amendola 173, 70126, Bari, Italy}

\maketitle

\abstracts{The possibility of performing post-Newtonian
gravitoelectromagnetic measurements with a joint LAGEOS-LAGEOS
II-OPTIS space-based mission is investigated}

OPTIS$^1$ is a recently proposed satellite--based mission which
would allow for precise tests of basic principles underlying
Special Relativity and post-Newtonian gravity. This mission is
based on the use of a spinning drag--free satellite in an
eccentric, high--altitude orbit which should allow to perform a
three orders of magnitude improved Michelson--Morley test and a
two orders of magnitude improved Kennedy--Thorndike test.
Moreover, it should also be possible to improve by two orders of
magnitude the tests of the universality of the gravitational
redshift by comparison of an atomic clock with an optical clock.
Since it is not particularly important for the present version of
the mission, the final orbital configuration of OPTIS has not yet
been fixed; in ref$^1$ a perigee height of 10000 km and apogee
height of 36000 km, with respect to Earth's surface, are
provisionally proposed assuming a launch with Ariane 5.

The requirements posed by the drag--free technology to be used,
based on the field emission electrical propulsion (FEEP) concept,
yield orbital altitudes not less than 1000 km. On the other hand,
the eccentricity $e$ should not be too high in order to prevent
passage in the Van Allen belts which could affect the on--board
capacitive reference sensor. Moreover, the orbital period $P_{\rm
OPT}$ should be shorter than the Earth's daily rotation of 24
hours. The orbital configuration proposed in ref$^1$ would imply a
semimajor axis $a_{\rm OPT}=29300$ km and an eccentricity $e_{\rm
OPT}=0.478$. With such values the difference of the gravitational
potential $U$, which is relevant for the gravitational redshift
test, would amount to ${\Delta U}/{c^2}\sim 1.8\times 10^{-10}$.
Such result is about three orders of magnitude better than that
obtainable in an Earth--based experiment.

An essential feature of OPTIS is the drag--free control of the
orbit. For a drag--free motion of the satellite a sensor measuring
the actual acceleration and thrusters counteracting any
acceleration to the required precision are needed. The sensor,
which is based on a capacitive determination of the position of a
test mass, has a sensitivity of up to$^2$ $10^{-12} \hbox{cm}\
{\hbox{s}}^{-2} {\rm Hz}^{-\rp{1}{2}}$.  Similar drag--free
systems of similar accuracy and with mission adapted modifications
will be used in MICROSCOPE, STEP and LISA. These systems have a
lifetime of many years.

In this paper we wish to investigate the possibility to use the
orbital data of OPTIS for performing precise tests of
post-Newtonian gravitoelectromagnetism as well$^3$. The
gravitomagnetic Lense-Thirring effect on the orbit of a test
body$^4$ is given by secular precessions of the longitude of the
ascending node $\Omega$ and the argument of pericentre $\omega$
\eqi \dot\Omega_{\rm LT} =\rp{2GJ}{c^2 a^3(1-e^2)^{{3}/{2}}},\
\dot\omega_{\rm LT} =-\rp{6GJ\cos i}{c^2 a^3(1-e^2)^{{3}/{2}}},
\eqf where $G$ is the Newtonian gravitational constant, $J$ is the
proper angular momentum of the central mass, $c$ is the speed of
light and $i$ is the inclination of the orbital plane to the
central mass's equator. The gravitoelectric pericentre advance
is$^{5}$ \eqi\dot\omega_{\rm GE}=\frac{3nGM}{c^2 a
(1-e^2)},\lb{perige}\eqf where $n=(GM/a^3)^{1/2}$ is the Keplerian
mean motion.

The rather free choice of the orbital parameters of OPTIS and the
use of a new drag--free technology open up the possibility to
extend its scientific significance with new important
post-Newtonian tests.
Indeed, it would be of great impact and scientific significance to
concentrate as many relativistic tests as possible in a single
mission, including also measurements in geodesy, geodynamics.
Another important point is that OPTIS is currently under serious
examination by a national space agency-the German DLR. Then, even
if it turns out that OPTIS would yield little or no advantages for
the measurement of the Lense--Thirring effect with respect, e.g.,
to the originally proposed LARES$^{6,7}$, if it will be finally
approved and launched it will nevertheless be a great chance for
detecting, among other things, the Lense-Thirring effect.

In Table 1 we report the orbital parameters of the existing or
proposed LAGEOS--type satellites and of the originally proposed
OPTIS configuration.
\begin{table}[htbp]
\ttbl{30pc}{Orbital parameters of \lg, \lgg, LARES and OPTIS}
{\begin{tabular}{lcccc}\\
\multicolumn{5}{c}{}\\[6pt]\hline
Orbital parameter & \lg & \lgg & LARES & OPTIS\\
\hline
$a$ (km) & 12270 & 12163 & 12270 & 29300\\
$e$ & 0.0045 & 0.014 & 0.04 & 0.478\\
$i$ (deg) & 110 & 52.65 & 70 & 63.4\\
$n$ (s$^{-1}$) & $4.643\times 10^{-4}$ & $4.710\times 10^{-4}$ &
$4.643\times 10^{-4}$ & $1.258\times 10^{-4}$\\
\hline
\end{tabular}}
\end{table}
\label{para}

The main characteristics of such a mission are the already
mentioned drag--free technique for OPTIS and the Satellite Laser
Ranging (SLR) technique for tracking. Today it is possible to
track satellites to an accuracy as low as a few mm. This may be
further improved in the next years.
With the level of accuracy reached with the most recent, although
preliminary, Earth gravity model solutions like GGM01C\footnote{It
can be retrieved on the WEB at
http://www.csr.utexas.edu/grace/gravity/. The GGM01C model
combines the Center for Space Research (CSR) TEG-4 model with data
from GRACE only. It seems to be very promising for our purposes.
Indeed, the released sigmas are not the mere formal errors but are
approximately calibrated.}, a three--nodes combination could be
considered. Indeed, by using the nodes of LAGEOS, LAGEOS II (with
a coefficient of $3\times 10^{-3}$) and OPTIS in the LARFES
orbital configuration (with a coefficient of 9.9$\times 10^{-1}$)
the relative error due to the static part of geopotential,
according to the variance matrix of GGM01C (RSS calculation) would
be $3\times 10^{-5}$, with a pessimistic upper bound of $6\times
10^{-5}$. The slope of the gravitomagnetic signal would be 61.4
mas yr$^{-1}$. In this case, since the nodes are insensitive to
the post--Newtonian gravitoelectric shift which, instead, affects
the perigee, the result of such test would be independent of the
inclusion of it into the force models. With the three--nodes
combination it should not be too optimistic to predict a total
error less than 1$\%$ over a time span of a few years.

The analysis of the perigee only of OPTIS-in the LARES orbital
configuration-could allow to measure the post-Newtonian
gravitoelectric shift. Indeed, for the LARES orbital configuration
\rfr{perige} yields a secular advance of 3280.1 mas yr$^{-1}$.
This implies that, for $\delta r^{\rm exp}\sim 1$ cm over 1 year,
the experimental error would be $\sim 1.3\times 10^{-3}$. The
impact of the mismodelling in the geopotential, according to the
present-day variance matrix of GGM01C, is of the order of $2\times
10^{-3}$, with a pessimistic upper bound of $4\times 10^{-3}$ due
to the sum of the absolute values of the individual errors. The
impact of $\delta (\dot J_2^{\rm eff})$ would yield a relative
error of $4\times 10^{-4}$ over one year. For the
non--gravitational perturbations, we could assume\footnote{E.g.
for 10-30 perigee perturbations with amplitudes of 0.2 mas
yr$^{-1}$. It can be shown using Table 4 of ref$^7$ that by using
only the perigee of LARES the non--gravitational perturbations
would have an impact of almost $4\times 10^{-3}$ over 7 years. }
an error of $\sim 6\times 10^{-4}-1\times 10^{-3}.$


\end{document}